\pdfoutput=1
\documentclass{sig-alternate-10pt}
\usepackage[margin=1in]{geometry}
\usepackage[T1]{fontenc}

\usepackage{xspace}
\usepackage{amsmath,amssymb}
\usepackage{xcolor}
\usepackage{hyperref}
\usepackage{amsmath,tabu}
\usepackage{hhline}
\usepackage{cite}
\usepackage{cleveref}
\usepackage{balance}

\usepackage{titlesec}

\newcommand{\dn}{D_\mathrm{n}}
\newcommand{\dx}{D_\mathrm{x}}
\newcommand\shapq{\mathit{Shapley}_\mathsf{Q}}
\newcommand\shapi{\mathit{Shapley}_\mathsf{I}}
\def\scs{\mathbf{S}}
\def\e#1{\emph{#1}}
\def\shapley{\mathit{Shapley}}
\newcommand{\pqe}{\mathrm{PQE}}
\def\set#1{\mathord{\{#1\}}}

\newcommand{\ignore}[1]{}

\newcommand{\defeq}{\mathrel{\mathop:}=}
\newcommand{\eqdef}{\mathrel{\mathop=}:}

\newcommand{\comlb}[1]{{\vspace{2mm}\noindent \bf  {\red{COMM(LEO):}}}~ #1 \hfill {\bf
    END.}\\}

\newcommand{\boxtheorem}{\hfill $\Box$}
\newcommand{\red}[1]{\textcolor{red}{#1}}

 \def\values{\mathbf{Val}}
\def\tup#1{\mathbf{#1}}
\def\dla{\mathbin{{:}{-}}}
\def\rel#1{\mathrm{#1}}
\def\ra{\rightarrow}
\def\PTIME{\textrm{PTime}\xspace}
\def\NP{\textrm{NP}\xspace}
\def\CONP{\textrm{coNP}\xspace}
\def\FP{\textrm{FP}\xspace}
\def\SHARPP{\textrm{\#P}\xspace}
\def\FPSHARPP{\textrm{FP$^{\textrm{\#P}}$}\xspace}
\def\Prob{\mathord{\mathrm{Pr}}}

\def\I{\mathcal{I}}

\def\calG{\mathcal{G}}

\def\MI{\mathsf{MI}}
\def\MC{\mathsf{MC}}

\def\Id{\I_{\mathsf{d}}}
\def\Imr{\I_{\mathsf{R}}}

\def\Imi{\I_{\MI}}
\def\Imc{\I_{\MC}}
\def\Ip{\I_{\mathsf{P}}}

\def\fpsharpp{\mathrm{FP}^{\mathrm{\#P}}}

\newenvironment{citedtheorem}[1]
{\begin{theorem}\hskip-0.2em\e{\cite{#1}}\,\,}
{\end{theorem}}

\newcommand{\algname}[1]{{\sf #1}}

\newtheorem{example}{Example}
\newtheorem{proposition}{Proposition}
\newtheorem{theorem}{Theorem}
\newtheorem{openpb}{Open Problem}

\title{The Shapley Value in Database Management}

\numberofauthors{2}

\author{
\alignauthor Leopoldo Bertossi\\
 \affaddr{SKEMA Business School}\\
 \affaddr{Montreal, Canada}\\
 \email{leopoldo.bertossi@skema.edu}
 \alignauthor Benny Kimelfeld\\
 \affaddr{Technion -- Israel Institute of Technology}\\
 \affaddr{Haifa, Israel}\\
 \email{bennyk@technion.ac.il}
 \and
\alignauthor Ester Livshits\\
 \affaddr{University of Edinburgh}\\
 \affaddr{Edinburgh, UK}\\
 \email{ester.livshits@ed.ac.uk}
 \alignauthor Mika\"{e}l Monet\\
 \affaddr{Université de Lille, CNRS, Inria, UMR 9189 - CRIStAL, F-59000 Lille}\\
 \affaddr{Lille, France}\\
 \email{mikael.monet@inria.fr}
}

\begin{document}
\maketitle
\begin{abstract}
Attribution scores can be applied in data management to quantify the contribution of individual items to conclusions from the data, as part of the explanation of what led to these conclusions. In Artificial Intelligence, Machine Learning, and Data Management, some of the common scores are deployments of the Shapley value, a formula for profit sharing in cooperative game theory. Since its invention in the 1950s, the Shapley value has been used for contribution measurement in many fields, from economics to law, with its latest researched applications in modern machine learning. Recent studies investigated the application of the Shapley value to database management. This article gives an overview of recent results on the computational complexity of the Shapley value for measuring the contribution of tuples to query answers and to the extent of inconsistency with respect to integrity constraints. More specifically, the article highlights lower and upper bounds on the complexity of calculating the Shapley value, either exactly or approximately, as well as solutions for realizing the calculation in practice.  
\end{abstract}

\section{Introduction}\label{sec:intro}

Explanations have been investigated
in Artificial Intelligence (AI) and Machine Learning (ML)
for some decades, and also, for a much longer time,
in other fields such as Philosophy, Logic, Physics, and Statistics. Actually, the explicit
study of explanations can be traced back to the ancient Greeks, who were already concerned
with causes and effects. A whole new area of research has emerged around explanations in AI ({\em Explainable AI} or \e{XAI} \cite{molnar}) and Data Science.
The basic need is fundamental: a human applies AI to make important decisions, and the human is accountable for these decisions as well as for the very choice to involve the specific AI algorithms in deployment. Hence, adopting the algorithm's decision may necessitate some level of human understanding of what led to the decision. 

In light of that, a decision or classification model, typically obtained through machine learning, should be able to attach an explanation to its outcome.  Stakeholders, such as an applicant for a loan from the bank, may desire a reason for the decision, especially if the request is declined. In a similar vein, a database may contain massive volumes of data or be built according to an intricate schema. Query answers could be difficult to explain or quantify in terms of the relevance of specific data in the database. This is also true for other database phenomena, such as the violation of integrity constraints.

In this article, we will concentrate mostly on explanations in data management, and more specifically, in relational databases (which we refer to simply as \e{databases} in the remainder of the manuscript). Different data-related phenomena may open a quest for explanations. Typically, one desires to understand \e{why} an answer is returned when applying the query to the database. One may also want to understand why a particular potential answer is \e{not} returned. In another scenario, they may wish to understand why an aggregate query result is a number that is unexpectedly high or low. It is also of interest to understand why an integrity constraint is not satisfied by the database. Even more, given a set of integrity constraints and a numerical measure of their violation by the database, we may want to know the extent to which a data item is responsible for the database's inconsistency~\cite{DBLP:conf/lpnmr/Bertossi19,DBLP:conf/sigmod/LivshitsKTIKR21, DBLP:journals/lmcs/LivshitsK22}.

We consider explanations that point to data items that contribute to the outcome that we wish to explain. As many items might take part in the outcome one way or another, we need to be able to \e{quantify} this contribution. To that aim, one of the fundamental concepts adopted is the \e{Shapley value}~\cite{shapley:book1952}, which is a formula for wealth distribution in a cooperative game. The Shapley value has a plethora of applications, including profit sharing between Internet providers~\cite{DBLP:journals/ton/MaCLMR10}, influence measurement in social networks~\cite{DBLP:journals/tase/NarayanamN11}, the importance of genes for specific body functions~\cite{moretti2007class}, and key-player identification in terrorist networks~\cite{DBLP:journals/snam/CampenHHL18}, to name a few. In the context of explanations, the idea is straightforward: data items are the players who play the game of establishing the outcome. 

It is relevant to emphasize that the Shapley value corresponds to a {\em principled approach} to quantify the contribution of a player to a wealth function that is shared by a set of players. Its inception started by stating some desired general properties  of such a contribution score. It was next proved that there is only one such score that satisfies those properties, namely the Shapley value \cite{roth}.

An inherent challenge in the application of the Shapley value for explanations (and other tasks) is its computational complexity---the Shapley value is often intractable to calculate, and particularly, the execution cost might grow exponentially with the number of players. Hence, past research  investigated islands of tractability and  approximation techniques for the Shapley value.

In the remainder of the introduction, we delve into some background on explanations in AI and their development into explanations in databases.

\ignore{
\comlb{I am making changes here. Up to here for now. I fixed a few things in the rest of the doc.}
}

\subsection*{Background on Explanations}

Explanations may come in different forms, and different kinds of them have been  proposed and investigated in the context of Computer Science. These include the area of {\em model-based diagnosis}, whose most prominent kinds are {\em consistency-based diagnosis} and {\em abductive diagnosis}~\cite{struss}.  \e{Causality}~\cite{pearl}, and particularly {\em actual causality}~\cite{HP05}, deem causes as explanations for observed effects and phenomena. In all these kinds of diagnosis, an explanation comes in the form of a set of basic propositions, which are expressed in the language of the model that describes the  situation related to the observations \cite{LeoRW22}. In the case of causality, they can be values of observed variables or features. 

Actual causality \cite{HP05,halpern} provides \e{counterfactual explanations} to observations. These explanations are obtained by hypothetically {\em intervening} in (i.e., changing some components  of)
the system under observation, to detect whether the intervention leads to changes in the observed behavior. 
In general terms, counterfactuals are basic propositions about components of a system that may be a cause for the observed behavior. 

Importantly, in actual causality, we distinguish between {\em endogenous} and {\em  exogenous} variables~\cite{HP05,Halpern2005-HALCAE-2}. Only the former are subject to interventions and may become actual causes. The latter are variables that we do not question, or have no control upon. The separation between endogenous and exogenous variables is application dependent.

Counterfactual causes are actual causes that
directly explain the observation: changing them leads to a change in the observation. Actual causes that are non-counterfactual ones are {\em weaker} causes, in that they require the company of other components to explain the observation. This idea is formalized in quantitative terms by means of the {\em responsibility score}~\cite{Chockler04} that captures the causal strength of an explanation. In this way, the score takes into account  the {\em amount} of company that a potential explanation requires in order to become a counterfactual cause. 
Accordingly, a counterfactual explanation
may come with a {\em responsibility score} that represents its causal strength in relation to the observation. A numerical score like this is usually called an {\em attribution score} (of the explanation for the observation). 

Attribution scores in the explanation of causes are widely investigated in ML. They are provided, most typically and frequently, for outcomes of ML-based decision and classification systems. 
 Next, we discuss a few prominent examples. (For a thorough description, c.f.~some recent surveys~\cite{Burkart,fosca,minh,molnar}.)

The {\em Resp} score~\cite{deem} applies the general responsibility score to the outcomes of classification systems. Specifically, it
quantifies the relevance of feature values in an input entity to an ML-based system for the obtained classification label. To this aim, Resp generalizes the responsibility score to deal with non-binary variables. The relevance of feature values is also quantified by the highly popular {\em SHAP} attribution score~\cite{{lund17,lundberg20}}.  SHAP is a particular application of the Shapley value. More precisely, SHAP \e{is} the Shapley value for a particular cooperative game played by a set of players that correspond to the features.
The tractability and approximability of SHAP for certain classes of classifiers has been recently investigated~\cite{AAAI21,arenas2021complexity,guyAAAI21}. The Resp and SHAP scores 
have also been experimentally compared~\cite{deem}.

Finally, in a different direction, the Shapley value was used to quantify the relevance of particular formulas to the inconsistency of a knowledge base~\cite{DBLP:journals/ai/HunterK10}.

\subsection*{Explanations  in Databases}\label{sec:xdb}

In database management, explanations usually come as database tuples or cells  that play a  role in the observed phenomenon, such as the received query answers. Actual causality and responsibility scores for query answers have been introduced and investigated~\cite{DBLP:journals/pvldb/MeliouGMS11,DBLP:journals/mst/BertossiS17,DBLP:journals/kais/Bertossi21}. In this context, connections between actual causality and model-based diagnosis have been established~\cite{DBLP:journals/mst/BertossiS17,DBLP:journals/ijar/BertossiS17}.
The counterfactual interventions are tuple updates: insertions or deletions of tuples, or changes of attribute values in them. These interventions are expected to change the outcome, for instance, to eliminate the query answer that we wish to explain. 
The identified tuples, as actual causes, are supplemented with responsibility scores, as additional quantitative information, that reflects their explanatory strength.  As in AI, we can partition the database into sets of \e{endogenous tuples} and \e{exogenous tuples}, and interventions are applied only to the former.

One can also formulate the causal approach to query-answer explanations in databases via {\em causal networks}~\cite{pearl}. For this purpose the {\em lineage} of the query~\cite{suciu} can be treated as a causal network. Actually, this has been the approach to define and compute another causality-based score for query answers, the {\em causal effect}, which has its roots in {\em causality for observational studies}~\cite{imbens}. The causal effect has been offered as an alternative to actual causality and responsibility in databases~\cite{tapp,lmcs}. The causal effect of a tuple $\tau \in D$ in relation to a Boolean query $q$, is  the expected difference $\mathbb{E}(q|\tau \in D) - \mathbb{E}(q|\tau \notin D)$, where the database is viewed a probabilistic database~\cite{suciu} (where every tuple can be eliminated randomly and independently) and
$q$ is treated as a random variable taking values $0$ or $1$. 

Close to the lineage of a query~\cite{DBLP:journals/vldb/BenjellounSHTW08}, we find the notion of {\em provenance} of a query~\cite{DBLP:journals/sigmod/BunemanT18,DBLP:conf/pods/GreenT17}, which can be used to explain a query answer by tracing back its origins to the underlying data source. 

More recently, Livshits et al.~\cite{lmcs} investigated the application of Shapley value to define explanations in databases. In this manuscript, we give a survey of this and other applications of the Shapley value as an explanation mechanism in databases. The first application we discuss
assigns to (endogenous) database tuples scores that reflect their importance for an obtained query answer. So, similarly to the application of the Shapley value in explainable machine learning, one has to define an appropriate coalitional game  that depends on the query and reflects this importance. (See \Cref{sec:query_answering}.) 

The Shapley value has 
also been applied to quantify the contribution of database tuples  to the inconsistency of the database in relation to {\em integrity constraints}, similarly to its application in knowledge-base inconsistency~\cite{DBLP:journals/ai/HunterK10}. This has been done for different measures of inconsistency~\cite{DBLP:journals/lmcs/LivshitsK22}. (See \Cref{sec:inconsistency}.)  In the same spirit, an attribution score for inconsistency that is directly based on {\em database repairs} \cite{DBLP:conf/pods/Bertossi19} has been introduced and investigated  \cite{DBLP:conf/lpnmr/Bertossi19}. 

\subsection*{Organization}
The remainder of the manuscript is organized as follows. We give preliminary teminology and concepts in \Cref{sec:preliminaries}. In \Cref{sec:query_answering}, we discuss the application of the Shapley value in the explanation of query answers, and in \Cref{sec:inconsistency} we do so for database inconsistency. Finally, we conclude in \Cref{sec:conclusions}.

\section{Basic Concepts}\label{sec:preliminaries}

We first present the basic concepts and main notation that we use throughout the article.

\paragraph{Databases.}
A \e{database schema} $\scs$ is a finite collection of \e{relation symbols} $R$, each with an associated signature $(A_1,\dots,A_\ell)$  of distinct \e{attributes} $A_i$. A \e{database} $D$ over a schema $\scs$ is a finite collection of \e{facts} of the form $R(c_1,\dots,c_k)$ where $R$ is a relation symbol of $\scs$ with the signature $(A_1,\dots,A_k)$, and each $c_i$ is a \e{value} that is in the domain of the attribute~$A_i$. For our complexity analysis, we assume that all database values come from a fixed infinite (recursively enumerable) domain $\values$; in particular, we assume that
the domain of every attribute is simply~$\values$.
A fact with the relation symbol $R$ is also called an \e{$R$-fact}.

\paragraph{Queries.}
Let $\scs$ be a schema. A \e{query} $q$ (over $\scs$) is associated with an arity $k\geq 0$ and it maps every database $D$ over $\scs$ into a finite $k$-ary relation of values (that is, a finite subset of $\values^k$). A query $q$ with arity zero is called a \e{Boolean} query, and then $q(D)$ is either \e{true} (i.e., consists of the empty tuple) or \e{false} (i.e., is empty). 

A special case of a query is a \e{Conjunctive Query} (CQ) that captures the select-project-join queries of SQL and has the logical form
$$\set{\tup{x} \mid \exists\tup{y}[\varphi_1(\tup{x},\tup{y})\land\cdots\land\varphi_m(\tup{x},\tup{y})]}$$
where $\tup{x}$ and $\tup{y}$ are disjoint sequences of distinct variables and each $\varphi_i(\tup{x},\tup{y})$ is an atomic formula over $\scs$, that is, a formula of the form $R(t_1,\dots,t_k)$ where $R\in\scs$ is  a $k$-ary relation symbol and each $t_i$ is either a value (constant) or a variable from $\tup x$ or from $\tup y$. The variables of $\tup x$ and $\tup y$ are called the \e{free} and the \e{existential} variables, respectively. 
For convenient notation, we denote a CQ $q$ by
$$q(\tup x)\dla \varphi_1(\tup{x},\tup{y}),\dots,\varphi_m(\tup{x},\tup{y})$$ as in the following example that finds all manager-employee pairs such that the manager manages the department of the employee:
$$\rel{Manages}(x,z) \dla \rel{EmpDept}(z,y),\rel{DeptMgr}(y,x)$$
We call $q(\tup x)$ the \e{head} of $q$ and 
$\varphi_1(\tup{x},\tup{y}),\dots,\varphi_m(\tup{x},\tup{y})$ the \e{body} of $q$. Each $\varphi_i(\tup{x},\tup{y})$ is an \e{atom} of $q$. A \e{self-join} of $q$ is a pair of atoms that use the same relation symbol. For example, the above 
$\rel{Manages}$ CQ is \e{self-join-free} since each relation symbol is used only once. 
Lastly, we recall that a \emph{union of conjunctive queries} (UCQ) is 
a logical disjunction of CQs with the same arity.

\paragraph{Integrity constraints.}
Let $\scs$ be a schema. An \e{integrity constraint} is simply a Boolean query. A database $D$ satisfies a set $\Delta$ of integrity constraints, denoted $D\models\Delta$, if $D$ satisfies every constraint in $\Delta$, that is, $D\models\gamma$ for all $\gamma\in\Delta$. When $\Delta$ is clear from the context, we say that $D$ is \e{consistent} if $D\models\Delta$, and otherwise that it is \e{inconsistent}. Two sets of integrity constraints are \e{equivalent} if every database that satisfies one also satisfies the other.

A special case of an integrity constraint is a \e{Functional Dependency} (FD) $R:X\ra Y$ where $R$ is a relation symbol of $\scs$ and $X$ and $Y$ are sets of attributes of $R$. A database $D$ satisfies $R:X\ra Y$ if every two $R$-facts that agree on (i.e., have the same value for every attribute of) $X$ also agree on $Y$. When $R$ is clear from the context, we may write just $X\ra Y$.

\paragraph{Shapley value.}
As explained in the introduction, the Shapley value is a function for wealth distribution in cooperative games. The precise definition is as follows.
Let $L$ be a finite set of players. A cooperative game is a function $\calG\colon \mathcal{P}(L) \to\mathbb{R}$, where $\mathcal{P}(L)$ is the power set of $L$. 
For $M\subseteq L$, the value $\calG(M)$ represents
a value, such as wealth, jointly obtained by $M$ when the players of $M$
cooperate. The Shapley value for the player $a$ in the game~$\calG$ is, intuitively, the following quantity. Suppose that we form a random cooperating team by selecting players, one by one, uniformly and without replacement; what is the expected change of utility when $a$ is selected? The exact definition is:
\begin{align}
  \shapley&(L, \calG, a) \defeq \label{eq:shapley}\\ \frac{1}{|L| !}  & \sum_{\pi \in \Pi_{L}} (\calG(\pi_{a} \cup \set{a}) - \calG(\pi_{a})) \notag 
  \end{align}
Here, $\Pi_{L}$ is the set of all possible permutations over the players in $L$, and for each permutation $\pi$ we denote by $\pi_a$ the set of players that appear before $a$ in the permutation.

\paragraph{Complexity concepts.} 
Throughout the paper, we refer to the standard classes \PTIME,
\NP and \CONP of decision problems. We will also consider the
function classes \FP, \SHARPP and \FPSHARPP. Recall that \FP is the
class of functions that can be computed in polynomial time, \SHARPP
is the class of functions can be described as counting the accepting
paths of a nondeterministic Turing machine on the given input, and
\FPSHARPP is the class of functions that can be computed in
polynomial time with an oracle to some function in \SHARPP.

We assume that integer numbers are represented in the usual binary way, and that non-integers are rational numbers represented using their numerator and denominator, where $(n,d)$ stands for $n/d$. 

Suppose that $f$ is a numerical function that maps its input $x$ to a number $f(x)$. A \e{Fully Polynomial-time Randomized Approximation Scheme} (FPRAS) for $f$ is a randomized algorithm $A$ that takes as input an instance $x$ of $f$ and an error $\epsilon>0$, and returns an $\epsilon$-approximation of $f(x)$ with probability at least $3/4$ (which is arbitrary and can be amplified to be arbitrarily close to $1$ using standard techniques). More precisely, we distinguish between an \e{additive FPRAS}:
$$\Prob\big[f(x)-\epsilon\leq A(x)\leq f(x)+\epsilon\big]\;>\; \frac34$$
and a \e{multiplicative FPRAS}:
$$\Prob\big[\frac{f(x)}{1+\epsilon}\leq A(x)\leq (1+\epsilon)f(x)\big]\;>\; \frac34$$

\section{Contribution to Database Queries}\label{sec:query_answering}

In this section, we survey the use of the Shapley value in the context of
explaining query answers. 
We will focus on
Boolean queries, to keep the presentation short,  but the framework easily extends to queries with free
variables, or to aggregate queries returning numerical values (e.g., as explained by Livshits et al.~\cite{lmcs}). 
We first define the notions and then review recent work.

As mentioned in the introduction, a database~$D$ consists
of a set~$\dx$ of \emph{exogenous} facts,
and a set~$\dn$ of \emph{endogenous} facts. 
For a Boolean query~$q$ and endogenous fact~$f\in \dn$ of a database~$D = \dn \cup \dx$, the goal is
to measure the contribution of~$f$ to the result $q(D)$. To this end, we view a Boolean query as a numerical query that answers $1$ if it is satisfied by the database and $0$ otherwise. We then apply the Shapley value, where the players are the endogenous facts and
the game function is the function $\calG_{q,\dx,\dn}:\mathcal{P}(\dn) \to \{0,1\}$ mapping every subset~$E\subseteq \dn$
to the value $q(E \cup \dx)-q(\dx)$. (The subtraction is applied to satisfy the requirement that the wealth function is zero on the empty set of players.) Applying Equation~\eqref{eq:shapley},
\emph{the Shapley value of $f$ (in $D$
for $q$)}, denoted~$\shapq(q,\dn,\dx,f)$, is defined as follows:
\begin{align*}
  \label{eq:shapleyq}
  & \shapq(q, \dn,\dx,f) \defeq\shapley(\dn,\calG_{q,\dx,\dn}, f) \\
 &=  \frac{1}{|\dn| !} \sum_{\pi \in \Pi_{\dn}} \big[q( \dx \cup \pi_{f} \cup \set{f})
- q(\dx \cup \pi_{f})\big].
\end{align*}

Intuitively, this value
represents the contribution of~$f$ to the result of the query: the higher this value
is, the more $f$ contributes to satisfying $q$.  Note that this
value could be negative, in the case where $q$ is not monotone.
Moreover, the properties of the Shapley value imply that we always
have $q(D) = q(\dx)  + \sum_{f\in \dn} \shapq(q, \dn,\dx,f)$. In
 words, the contributions of all endogenous facts
sum
up to $q(D)-q(\dx)$; hence, the Shapley value states how the score $q(D)$ on the whole database is to be 
shared among the endogenous facts.

\subsection{Exact Computation}

In database theory, the computational complexity of a problem is
often measured with what is called the \emph{data
complexity}~\cite{vardi1982complexity}, which is the complexity of
the problem when a particular query~$q$ is fixed (there is, thus,
one computational problem for each distinct query).
Accordingly, past work~\cite{lmcs,DBLP:conf/pods/ReshefKL20,deutch2022computing}
has studied the data complexity of exactly computing the Shapley
value of a fact, as defined above, depending on the particular
query $q$ at hand. Formally, for an arbitrary Boolean query $q$ and slightly abusing the notation, the computational problem $\shapq(q)$ is defined as follows:
\begin{center}
\fbox{\begin{tabular}{rp{5.3cm}}
\sc{Problem}: & $\shapq(q)$ 
\\
\sc{Param}: & Boolean query $q$ 
\\
\sc{Input}: & Database $\dx \cup \dn$, fact $f \in \dn$
\\
\sc{Goal}: & Compute $\shapq(q,\dn,\dx,f)$
\end{tabular}}
\end{center}
We aim for dichotomies in complexity, for 
classes of queries (e.g., CQs), that chart the boundary between polynomial-time cases and intractable cases. We report on such results in this section.

\subsubsection{Reduction to Probabilistic Databases}

Deutch et al.~\cite{deutch2022computing} showed that the calculation of the Shapley value can be reduced to \emph{query answering in probabilistic databases}. Hence, this reduction yields a class of tractable queries. To explain that, we need some definitions.

A \emph{Tuple-Independent Database} 
(\e{TID} for short) is a pair $(D,\pi)$ consisting of a
database $D$ and a function $\pi$ that maps each fact $f \in D$ to a
probability $\pi(f) \in [0,1]$. The TID $(D, \pi)$ defines a
probability distribution $\Pr_{D,\pi}$ on $\mathcal{P}(D)$, where
each $D' \subseteq D$ has the probability
$\Pr_{D,\pi}(D') \eqdef \prod_{f \in D'} \pi(f) \times \prod_{f \in D
\backslash D'} (1 - \pi(f))$. We evaluate a Boolean query $q$ by calculating the probability that $q$ is satisfied by the TID: 
$\Pr(q, (D,
\pi)) = \sum_{D' \subseteq D}
\Pr_{D,\pi}(D')\cdot q(D')$. 
The problem of
\emph{Probabilistic Query Evaluation} for
$q$, or $\pqe(q)$ for short, is the following:
\begin{center}
\fbox{\begin{tabular}{rp{5.2cm}}
\sc{Problem}: & $\pqe(q)$ 
\\
\sc{Param}: & Boolean query $q$ 
\\
\sc{Input}: & TID $(D,\pi)$
\\
\sc{Goal}: & Compute $\Pr(q, (D, \pi))$
\end{tabular}}
\end{center}

We can now state the result of Deutch et al.~\cite{deutch2022computing}:

\begin{theorem}
\label{thm:to-pqe}
For every Boolean query $q$, the problem $\shapq(q)$ reduces in
polynomial time to the problem $\pqe(q)$.
\end{theorem}

Notice that this result is quite general, in that~$q$ can be an
arbitrary Boolean query: a CQ, an FO query, an MSO, Datalog or RPQ
query---it does not matter: if $\pqe(q)$ is tractable then $\shapq(q)$ is tractable as well.  As it turns, a celebrated result by Dalvi and
Suciu~\cite{dalvi2013dichotomy} provides a dichotomy on unions of
conjunctive queries 
for $\pqe$: either~$q$ is \emph{safe} and $\pqe(q)$ is solvable in
polynomial time (\FP), or $q$ is not safe and $\pqe(q)$ is
$\FPSHARPP$-hard. Therefore, a direct corollary of \Cref{thm:to-pqe} is
that $\shapq(q)$ is in \FP for all safe UCQs.

\subsubsection{Calculation via Knowledge Compilation}

Deutch et al.~\cite{deutch2022computing} also provide another
route for solving this problem in practice, through \emph{knowledge compilation}. We only sketch this approach here and
refer to their work for more details. The idea is to first
compute, for the query~$q$ and database~$D$, the \emph{lineage} (also called \emph{provenance})
of~$q$ on~$D$, which is a Boolean circuit that intuitively captures
the dependence of the query answer on individual facts of~$D$.
This lineage is then transformed, using a \emph{knowledge
compiler} tool, into an equivalent Boolean circuit in restricted classes 
from knowledge compilation (namely, so-called \emph{deterministic and decomposable Boolean
circuits}), over which the authors of~\cite{deutch2022computing} design a polynomial-time algorithm to
compute the Shapley values. 

The queries that can be solved
in \FP with this method are a subset\footnote{And it is
unknown if this is a strict subset or not, see, e.g., \cite{monet2020solving}.} of those that are
captured with \Cref{thm:to-pqe}, so in terms of
theoretical results nothing is gained here (apart from a polynomial of
slightly lower degree). Nevertheless, this method allows them to
use existing tools to solve the problem in practice instead of
implementing everything from scratch. They used in particular 
ProvSQL~\cite{senellart2018provsql}, a tool integrated
into PostgreSQL that can perform lineage computation in various
semirings, and the knowledge compiler
c2d~\cite{darwiche2004new}. 

\subsubsection{Hardness and Dichotomy}

As we have discussed, \Cref{thm:to-pqe} allows to
capture a large class of tractable queries for the
problem~$\shapq$. It is, however, still unknown whether this actually captures all tractable cases. In particular, it is unknown whether there also exists a general reduction in the other direction:
\begin{openpb}
	\label{open:pqe-to-shap}
	Does $\pqe(q)$ reduce in polynomial time to $\shapq(q)$ for all queries $q$?
\end{openpb}
Combined with known results on the complexity of probabilistic databases, a
positive answer to this question would then yield a complete
dichotomy of $\shapq$ for the class of all UCQs and, in fact, even for the more general
class of queries \emph{closed under
homomorphism}~\cite{DBLP:conf/icdt/Amarilli23}.

For the class of CQs without self-joins, Livshits et al.~\cite{lmcs}  
obtained a dichotomy. The tractability condition is that $q$ is \e{hierarchical}: for all variables $y$ and $y'$ it holds that
$A_y\subseteq A_{y'}$, or $A_{y'}\subseteq A_y$, or
$A_y\cap A_{y'}=\emptyset$, where $A_x$ is the set of atoms of $q$ that use the variable $x$~\cite{DBLP:journals/cacm/DalviRS09}. 
\begin{theorem}
\label{thm:sjf-dicho}
Let $q$ be a self-join-free CQ. If~$q$ is hierarchical, then
$\shapq(q)$ is in \FP, otherwise it is
$\FPSHARPP$-hard.
\end{theorem}

We conclude this part with some comments on \Cref{thm:sjf-dicho}.
First, the positive part is already captured by
\Cref{thm:to-pqe}, as it is known that hierarchical CQs
are safe for PQE on TIDs~\cite{dalvi2013dichotomy}. In contrast, the lower bound 
requires a reduction that is specifically crafted for the problem. In particular, it is not clear whether and how this reduction can be generalized to solve 
\Cref{open:pqe-to-shap},  and there is no generalization of \Cref{thm:sjf-dicho} for CQs with self-joins and UCQs. Second, the reduction for the lower bound of \Cref{thm:sjf-dicho} requires the usage of both endogenous and exogenous facts, and it is still open whether the lower bound holds if we assume that all facts are endogenous. 
Finally, Reshef et al.~\cite{DBLP:conf/pods/ReshefKL20} studied the
complexity of $\shapq$ for CQs with negated atoms (but still no self-joins), and established analogous dichotomy results under the restrictions that certain relations can contain only exogenous
facts.

\subsection{Approximation}

As we have seen, computing the exact Shapley values of facts is
not always tractable in theory. This naturally brings the question
of which queries allow for approximate computation of these
values with desired guarantees on the data complexity and the approximation ratio. We review in this section what
is known on the approximability of the problems~$\shapq(q)$, in
terms of FPRAS. 

As pointed out in~\cite{lmcs}, by using the
Chernoff-Hoeffding bound one can easily obtain an additive FPRAS for computing
$\shapq(q,\dn,\dx,f)$: this can be done for instance by sampling
$O(\log(1/\delta)/\epsilon^2)$ permutations $\pi$ of~$\dn$ and computing the
average value of $q( \dx \cup \pi_{f} \cup \set{f}) - q(\dx \cup \pi_{f})$,
assuming that $q$ itself is answerable in polynomial time. Note that this assumption is
true for the most common query classes such as UCQs, Datalog, etc.
Formally:
\begin{proposition}
  \label{prp:add-fpras}
  If $q$ can be answered in polynomial time, then 
  $\shapq(q)$ has an additive FPRAS.
\end{proposition}

An additive FPRAS might be insufficient when the values to approximate are very small. Nevertheless, for a large
class of queries, the Shapley values cannot in fact be too small~\cite{lmcs}.
Specifically, a (Boolean) query $q$ is said to have the \emph{gap
property} if the Shapley value of any fact is either zero or is
``not too small,'' that is, at least the reciprocal of a polynomial. Then, for a query with polynomial-time evaluation \e{and} the gap property, the additive FPRAS from \Cref{prp:add-fpras} can be easily transformed into a
multiplicative FPRAS. As it turns out, all UCQs have the gap property~\cite[Proposition 4.12]{lmcs}. 
\begin{proposition}
  \label{prp:mult-fpras}
  If $q$ has polynomial-time evaluation and the gap
  property, then $\shapq(q)$ admits a multiplicative FPRAS. This
  holds for all UCQs.
\end{proposition}

Not all queries satisfy the gap property. For instance, some CQs
with negated atoms (that we denote by CQ$^\lnot$) may lack this property. In fact,
by establishing a connection between the existence of a
multiplicative FPRAS for~$\shapq(q)$ and what is called the
\emph{relevance problem for~$q$}, Reshef et al.~\cite{DBLP:conf/pods/ReshefKL20} were able to exhibit a
CQ$^\lnot$ that does not admit any multiplicative FPRAS (unless $\mbox{P}=\mbox{RP}$, which is widely believed to be false). So far, however, no dichotomy is known on the existence of a multiplicative FPRAS for CQ$^\lnot$s  (or even self-join--free CQ$^\lnot$s). In fact, we do not know any example of a
CQ$^\lnot$ that does not have the gap property but  nevertheless admits a
multiplicative FPRAS.

\subsection{Remarks}
We conclude with several remarks on the Shapley value for database queries. 
Interestingly, the state of affairs for approximate Shapley computation contrasts with what happens with the SHAP score in
machine learning. Arenas et al.~\cite{arenas2021complexity} have recently shown that the SHAP score does not admit a multiplicative FPRAS (under conventional assumptions in complexity theory), even for very simple monotone Boolean models (namely, monotone 2-DNFs)~\cite[Theorem 9]{arenas2021complexity}.

A possible alternative to the computation of the Shapley value is to \emph{rank} the endogenous facts according to their Shapley value, while possibly avoiding the actual computation of these values.  Indeed,
as far as we know, there could exist queries $q$ such
that $\shapq(q)$ is hard to approximate while the ranking problem is tractable, or vice-versa.   
Deutch et al.~\cite{deutch2022computing} proposed a fast heuristic to solve this problem, and this heuristic seems to work well empirically. 
Arad et al.~\cite{arad2022learnshapley} employed machine learning to learn this ranking.
Yet, as of today, a formal study of the complexity of the ranking problem is lacking. For instance, we do not know whether there are queries with intractable Shapley but tractable ranking.  Again, this
contrasts with what is known on the SHAP-score from machine
learning, as it has been shown that the ranking problem for this
score is unlikely to be in BPP even for simple monotone Boolean
models; see \cite[Theorem 16]{arenas2021complexity}.

Finally, Khalil and Kimelfeld~\cite{khalil2022complexity} have recently studied the complexity of the Shapley value in the context of graphs with labeled edges. Their problem is similar to what we discussed in this section, except that the players are the (endogenous) edges and/or vertices of the graph, and the queries are Regular Path Queries (RPQs) and Conjunctions of RPQs (CRPQs). An RPQ is associated with a regular expression $\gamma$, and a pair $(u,v)$ of vertices is an answer if any path from $u$ to $v$ conforms to $\gamma$. They established dichotomies for exact and approximate calculation of the Shapley value. For example, under conventional complexity assumptions, an RPQ has a multiplicative FPRAS if and only if the regular expression recognizes a finite language.

\section{Contribution to Database Inconsistency}\label{sec:inconsistency}

It is often wrong to assume that the database is clean of errors and conforms to our integrity constraints. 
Data might be the result of integrating unreliable sources (e.g., social media) that  contain mistakes and conflicting information. Moreover, the content of the database may be produced by error-prone procedures (e.g., natural-language or image processing). Given this nature of data, \e{measures of database inconsistency} quantify the extent to which the database violates a given set of integrity constraints~\cite{DBLP:conf/lpnmr/Bertossi19,DBLP:conf/sigmod/LivshitsKTIKR21, DBLP:journals/lmcs/LivshitsK22}. Inconsistency measures can be used, for example, to estimate the reliability of new datasets~\cite{DBLP:conf/atal/CholvyPRT15} or to build progress indicators for data-cleaning systems~\cite{DBLP:conf/sigmod/LivshitsKTIKR21}. Moreover, such measures can be used to attribute to database facts a level of responsibility to inconsistency, and so to prioritize facts in the explanation, inspection, or resolution of database inconsistency; this task is what we discuss in this part.

The attribution of responsibility to the inconsistency of the database relies on two main components: \e{(1)} an inconsistency measure \e{and (2)} a responsibility-sharing mechanism. For the former, we discuss several alternatives in the next section. For the latter, we use the Shapley value.

\subsection{Inconsistency Measures}\label{sec:inconsistency_measures}

The measurement of the inconsistency of information has been extensively studied by the Knowledge Representation (KR) and Logic communities~\cite{DBLP:conf/ijcai/KoniecznyLM03,DBLP:conf/kr/HunterK06,DBLP:journals/jiis/GrantH06,DBLP:journals/ai/HunterK10,DBLP:journals/ki/Thimm17}, where several different measures have been introduced. Some of these have been adapted to the database setting~\cite{DBLP:journals/lmcs/LivshitsK22}.

In general, an \e{inconsistency measure} $\I$ is a function that maps pairs $(D,\Delta)$ of a database $D$ and a set $\Delta$ of integrity constraints (ICs) to numbers $\I(D,\Delta)\in [0,\infty)$. 
Intuitively, the higher $\I(D,\Delta)$ is, the stronger $D$ violates $\Delta$. We make only the (reasonable) assumption that $\I(D,\Delta)$ is zero whenever $D$ is empty.
Here again, we use the Shapley value framework, where this time the
set of players is the whole database, and the utility is the
function 
$\calG_{D,\Delta, \I}:\mathcal{P}(D) \to \{0,1\}$ that maps every subset $D'\subseteq D$
to the value $\I(D',\Delta)$. Hence, applying Equation~\eqref{eq:shapley},
\emph{the Shapley value of a fact $f$ of $D$ w.r.t.~$\Delta$}, denoted~$\shapi(D, \Delta, \I, f)$, is defined as
\begin{align*}
 & \shapi(D, \Delta, \I, f) \defeq \shapley(D,\calG_{D,\Delta, \I}, f) \\
 &=  \frac{1}{|D| !} \sum_{\pi \in \Pi_{D}} (\I(\pi_{f} \cup \set{f},\Delta) - \I(\pi_{f},\Delta))\,.
\end{align*}
Following Livshits and Kimelfeld~\cite{DBLP:journals/lmcs/LivshitsK22}, here we do not distinguish between exogenous and endogenous facts---all facts are considered endogenous (and all definitions naturally extend to account for exogenous facts). 

Livshits and Kimelfeld~\cite{DBLP:journals/lmcs/LivshitsK22} studied the Shapley value for several inconsistency measures that were previously explored in the context of databases and knowledge bases. (See~\cite{DBLP:conf/sigmod/LivshitsKTIKR21} for a study of the behavior of these measures from both a practical and a theoretical perspective.) In the following definitions, $D$ denotes a database and $\Delta$ a set of ICs.
\begin{enumerate}
    \item $\Id$, called the \e{drastic measure}~\cite{DBLP:journals/ki/Thimm17}  takes the value $1$ if the database is inconsistent and the value $0$ otherwise. 
    \item $\Imi$ counts the \e{minimal inconsistent subsets} of the database~\cite{DBLP:conf/kr/HunterK08,DBLP:journals/ai/HunterK10}.
    In notation, $\Imi(D,\Delta)\defeq|\MI(D,\Delta)|$ where
    $\MI(D,\Delta)$ is the set of all inconsistent subsets $E\subseteq D$ such that $E'\models\Delta$ for all $E'\subset E$. 
    \item $\Ip$ counts the \e{problematic facts}, where a fact is \e{problematic} if it belongs to a minimal inconsistent
      subset~\cite{DBLP:conf/ecsqaru/GrantH11}. In notation,
      $\Ip(D,\Delta) \defeq |\bigcup_{E \in \MI(D,\Delta)} E|$.
    \item $\Imr$ is the minimal number of facts that should be deleted from the database in order to satisfy      $\Delta$~\cite{DBLP:conf/ecsqaru/GrantH13,DBLP:journals/jacm/GoldreichGR98,DBLP:conf/lpnmr/Bertossi19}.
      In notation, 
      $\Imr(D,\Delta)\defeq \min_{E\subseteq D,D\setminus E\models\Delta}(|E|)$.
    \item $\Imc$ counts the \e{maximal consistent subsets}~\cite{DBLP:conf/ecsqaru/GrantH11,DBLP:journals/ijar/GrantH17} (also called \e{subset repairs}~\cite{DBLP:conf/pods/ArenasBC99}).
      In notation, $\Imc(D,\Delta)\defeq |\MC(D,\Delta)|$ where
      $\MC(D,\Delta)$ is the set of all consistent subsets $E\subseteq D$ such that $E'\not\models\Delta$ whenever $E\subset E'$. 
\end{enumerate}

Intuitively, the measure $I_d$ is simply an indicator of inconsistency. The measure $\Imi$ counts the violations (i.e.,~the minimal sets of facts that jointly violate the constraints) and the measure $\Ip$ counts the facts involved in such violations (and a fact is counted once even if it occurs in multiple violations). For both measures, the higher the number is, the more the constraints are violated; hence, the more inconsistent the database is.
The measure $\Imr$ quantifies the distance of the database from a consistent one---the more facts we have to remove to obtain consistency, the higher the measure is. Finally, the measure $\Imc$ counts the subset repairs; that is, all the different ways to obtain a consistent database by deleting a minimal set of facts---the more repairs there are, the more inconsistent the database is.

\subsection{Complexity of Exact Computation}

  \begin{table*}[t]
  \renewcommand{\arraystretch}{1.2}
  \centering
  \caption{The complexity of the (exact ; approximate) Shapley value of different inconsistency measures.
    \label{table:complexity}}
\begin{tabular}{c||c|c|c}
        & \textbf{lhs chain} & \textbf{no lhs chain, PTime cardinality repair} & \textbf{other}\\\hline\hline
        $\Id$ & PTime & \multicolumn{2}{c}{$\fpsharpp$-complete ;  FPRAS} \\ \hline
        $\Imi$ & \multicolumn{3}{c}{ PTime } \\ \hline
        $\Ip$ & \multicolumn{3}{c}{ PTime } \\ \hline
        $\Imr$ & PTime & \textbf{?} ; FPRAS &
                             NP-hard~\cite{DBLP:journals/tods/LivshitsKR20} ; no FPRAS \\ \hline
        $\Imc$ & PTime & \multicolumn{2}{c}{$\fpsharpp$-complete~\cite{DBLP:conf/pods/LivshitsK17} ; \textbf{?}} \\ \hline
    \end{tabular}
  \end{table*}

We proceed to discuss the complexity of computing the value $\shapi(D,\Delta,\I,f)$ for each one of the aforementioned inconsistency measures $\I$, as studied by Livshits and Kimelfeld~\cite{DBLP:journals/lmcs/LivshitsK22}. Their study considered the case where $\Delta$ is a set of FDs. We consider again the data complexity of computing the Shapley value; this time, the schema, set of FDs, and inconsistency measure are considered fixed, and the input consists of a database $D$ and a fact $f$. Accordingly, and again slightly abusing notation, we denote $\shapi(\Delta,\I)$
the corresponding computational problems.
\begin{center}
\fbox{\begin{tabular}{rp{5.2cm}}
\sc{Problem}: & $\shapi(\Delta,\I)$ 
\\
\sc{Param}: & Set $\Delta$ of ICs, inconsistency m.~$\I$ 
\\
\sc{Input}: & Database $D$ and fact $f \in D$ 
\\
\sc{Goal}: & Compute $\shapi(D,\Delta,\I,f)$
\end{tabular}}
\end{center}

It turns our that each of the inconsistency measures entails quite a unique picture of complexity.

\def\mparagraph#1{\medskip\par\noindent
\underline{\textbf{The measure #1}.}}

\mparagraph{$\Id$} While the drastic measure is the simplest one conceptually, it might be intractable to compute the Shapley value of a fact w.r.t.~$\Id$ even for simple FD sets. Livshits and Kimelfeld~\cite{DBLP:journals/lmcs/LivshitsK22} established a dichotomy in data complexity for this measure, based on the definition of left-hand-side chain~\cite{DBLP:conf/pods/LivshitsK17}. An FD set $\Delta$ has a left-hand-side chain (lhs chain, for short) if for every two FDs $X\rightarrow Y$ and $X'\rightarrow Y'$ in $\Delta$, either $X\subseteq X'$ or $X'\subseteq X$.

\def\theoremdrastic{
Let $\Delta$ be a set of FDs. If $\Delta$ is equivalent to an FD
set with an lhs chain, then $\shapi(\Delta,\Id)$ is in~$\FP$.
Otherwise, the problem is $\fpsharpp$-complete.}
\begin{theorem}\label{thm:drastic}
  \theoremdrastic
\end{theorem}

Theorem~\ref{thm:drastic} implies, for example, that the Shapley value can be computed efficiently for the set $\{A\ra B,AC\ra D\}$, but not for  $\{A\ra B,B\ra A\}$. As a proof technique, they show hardness directly only for $\{A\ra B,B\ra A\}$. For any other intractable FD set, hardness is established via the so called \e{fact-wise reductions}~\cite{DBLP:conf/pods/KimelfeldVW11}. Such reductions essentially map databases over one schema and set of ICs to databases over another schema and set of ICs while preserving (in)consistency. The positive side of Theorem~\ref{thm:drastic} is established via dynamic programming.

We note that the tractability criterion of Theorem~\ref{thm:drastic}  is the same as the tractability criterion for the problem of counting subset repairs (or, equivalently, maximal consistent subsets), and is decidable in polynomial time~\cite{DBLP:conf/pods/LivshitsK17}. 

\mparagraph{$\Imi$}
For a set $\Delta$ of FDs, it is easy to see that $\Imi$ simply counts the pairs of facts of the database that jointly violate the FDs. Hence, for a fact $f$ and some permutation $\pi$, we have that $\Imi(\pi_{f} \cup \set{f},\Delta) - \Imi(\pi_{f},\Delta)$ is precisely the number of facts in $\pi_f$ that are in conflict with~$f$. This simple observation implies the tractability of  $\Imi$.

\begin{citedtheorem}{DBLP:journals/lmcs/LivshitsK22}\label{thm:imi}
$\shapi(\Delta,\Imi)$ is in \FP for every set $\Delta$ of FDs.
\end{citedtheorem}

\mparagraph{$\Ip$} For this measure, we have that $\Ip(\pi_{f} \cup \set{f},\Delta) - \Ip(\pi_{f},\Delta)$ is the number of facts in $\pi_f$ that: \e{(1)} are in conflict with~$f$, and \e{(2)} are not in conflict with any other fact of $\pi_f$ (as, otherwise, they are considered ``problematic'' already before adding the fact $f$ to the picture). Based on this observation, we conclude the following.

\begin{citedtheorem}{DBLP:journals/lmcs/LivshitsK22}\label{thm:ip}
$\shapi(\Delta,\Ip)$ is in \FP for every set $\Delta$ of FDs.
\end{citedtheorem}

\mparagraph{$\Imr$}
The measure $\Imr$ is the only one for which we lack the full complexity picture; however, it is known that there are substantial sets of FDs for which this measure is tractable and others for which it is intractable. The hard cases follow from a prior work on the related problem of computing the cost of a \e{cardinality repair} (i.e., the minimal number of facts to remove to obtain consistency), as we explain next. 

Livshits et al.~\cite{DBLP:journals/tods/LivshitsKR20} established a dichotomy for finding a cardinality repair, where the tractability criterion is based on a polynomial-time algorithm, \algname{Simplify}, that simplifies the FD set in an iterative manner until no further simplification can be applied. If the result of \algname{Simplify}$(\Delta)$ is an empty FD set, then the problem is solvable in polynomial time; otherwise, it is NP-hard. Now, one of the basic properties of the Shapley value is ``efficiency''---the sum of the Shapley values over all the players equals the total wealth~\cite{shapley:book1952}. This property implies that for an inconsistency measure $\I$, it holds that
$$\sum_{f\in D}\shapi(D,\Delta,\I,f)=\I(D,\Delta)\,.$$
Thus, whenever the measure itself is hard to compute, so is the Shapley value of facts. This leads to the following result.

\begin{theorem}\label{thm:imr_hard}
Let $\Delta$ be a set of FDs. If the procedure \algname{Simplify}$(\Delta)$ of Livshits et al.~\cite{DBLP:journals/tods/LivshitsKR20} returns a nonempty set, then $\shapi(\Delta,\Imr)$ is NP-hard.
\end{theorem}

As in the case of the drastic measure, it can be shown, via a dynamic programming algorithm, that for an FD set with an lhs chain, the measure $\Imr$ is tractable.

\begin{theorem}\label{thm:imr_poly}
Let $\Delta$ be a set of FDs. If $\Delta$ is equivalent to an FD set with an lhs chain, then $\shapi(\Delta,\Imr)$ is in \FP.
\end{theorem}

It remains unknown what is the complexity of the problem for FD sets that fall outside the two cases covered by Theorems~\ref{thm:imr_hard} and~\ref{thm:imr_poly}.
\begin{openpb}
	\label{open:shapi}
What is the data complexity of $\shapi(\Delta,\Imr)$ for any set 
 $\Delta$ of FDs such that $\Delta$ has no lhs chain (up to equivalence) and \algname{Simplify}$(\Delta)$ returns an empty set? 
\end{openpb}
In particular, the problem is open for the FD set $\{A\ra B,B\ra A\}$ over the relation symbol $R(A,B)$.

\mparagraph{$\Imc$} As we explained for the previous measure, the efficiency property of the Shapley value allows us to conclude that the problem $\shapi(\Delta,\Imc)$ is hard whenever it is hard to calculate $\Imc(D,\Delta)$. Livshits et al.~\cite{DBLP:journals/jcss/LivshitsKW21} have shown that the subset repairs (i.e.,~maximal consistent subsets) can be counted in polynomial time for FD sets with an lhs chain (up to equivalence), and is \#P-complete for any other FD set. Hence, an lhs chain is a necessary condition for
tractability. As in the case of the measures $\Id$ and $\Imr$, it can be shown that this is also a sufficient condition using a dynamic programming algorithm. Hence, the following holds.

\def\theoremcounting{
Let $\Delta$ be a set of FDs. If $\Delta$ is equivalent to an FD set with an lhs chain, then $\shapi(\Delta,\Imc)$ is in \FP. Otherwise, the problem is $\fpsharpp$-complete.}

\begin{citedtheorem}{DBLP:journals/lmcs/LivshitsK22}\label{thm:counting}
    \theoremcounting
\end{citedtheorem}

The complexity results of this section are summarized in \Cref{table:complexity}. (The results on approximation are discussed in the next section.)

\subsection{Complexity of Approximation}
We have seen that, as far as exact computation is concerned, $\Imi$ and $\Ip$ are tractable for every set of FDs, but the measures $\Id$, $\Imr$, and $\Imc$ can be intractable even for simple sets of FDs. Therefore, we now discuss the approximate computation of $\shapi$. Again, the complexity results are of Livshits and Kimelfeld~\cite{DBLP:journals/lmcs/LivshitsK22}.

As in the case of $\shapq$, using the Chernoff-Hoeffding bound and a Monte-Carlo approach, one can easily obtain an additive FPRAS for computing $\shapi(D,\Delta,\I,f)$,
assuming that $\I(D,\Delta)$ can be computed in polynomial time, given $D$. A multiplicative FPRAS can be obtained using the same
  procedure 
  if a corresponding ``gap'' property holds, meaning that when the Shapley value is not zero, it is guaranteed to be ``large enough''.

  For instance, in the case of  the drastic measure, it can be shown that for
  all databases $D$ and facts $f$, $\shapi(D,\Delta,\Id,f)$ is either zero or at least $\frac{1}{|D|\cdot(|D|-1)}$.
Since $\Id(D,\Delta)$ can be computed in polynomial time for every $\Delta$, we conclude that:
\begin{proposition}\label{cor:drastic-approx}
 $\shapi(\Delta,\Id)$ has an additive and a multiplicative FPRAS for every set $\Delta$ of FDs.
\end{proposition}

Regarding $\Imr$, when it can be computed in polynomial time (i.e., when $\algname{Simplify}(\Delta)=\emptyset$), we can similarly show that $\shapi(\Delta,\Imr)$ has both an additive and a multiplicative FPRAS. The gap property holds here as well: $\shapi(D,\Delta,\Imr,f)=0$ or $\shapi(D,\Delta,\Imr,f)\ge \frac{1}{|D|\cdot(|D|-1)}$. In contrast, when $\Imr$ is intractable, it is not only NP-hard, but also APX-complete.
Consequently, there exists a polynomial-time constant-ratio approximation for $\Imr(D,\Delta)$, but for some $\epsilon>1$ there is no (randomized) $\epsilon$-approximation or else 
$\mbox{NP}=\mbox{RP}$. 
Then, the fact that $\shapi(D,\Delta,\Imr,f)\ge 0$ implies that the existence of a multiplicative FPRAS for the value $\shapi(D,\Delta,\Imr,f)$ would imply the existence of a multiplicative FPRAS for $\Imr(D,\Delta)$ via the efficiency property of the Shapley value. Therefore, the following holds.
\begin{theorem}\label{thm:mr-fpras}
  Let $\Delta$ be a set of FDs.
\begin{enumerate} 
\item If $\algname{Simplify}(\Delta)=\emptyset$, then $\shapi(\Delta,\Imr)$ has both an additive and a multiplicative FPRAS. 
\item Otherwise, $\shapi(\Delta,\Imr)$  has neither a multiplicative nor an additive FPRAS, assuming 
  $\mbox{RP}\neq\mbox{NP}$.
\end{enumerate}
\end{theorem}
Interestingly, it is still unknown whether there is \e{any} polynomial-time constant-ratio multiplicative approximation for $\shapi(D,\Delta,\Imr,f)$ for any FD set in the intractable side of \Cref{thm:mr-fpras}.

\begin{openpb}
	\label{open:constant-approx-Imr}
For any set $\Delta$ of FDs such that $\algname{Simplify}(\Delta)$ is nonempty, is there a polynomial-time constant-ratio multiplicative approximation for $\shapi(D,\Delta,\Imr,f)$?
\end{openpb}

Finally, the complexity picture for the approximate computation of the Shapley value w.r.t.~$\Imc$ is rather incomplete.
This is because counting the maximal consistent subsets w.r.t.~$\{A\rightarrow B,B\rightarrow A\}$ over $R(A,B)$ is the same as
counting the maximal matchings of a bipartite graph. As
the values $\shapi(D,\Delta,\Imc,f)$ are nonnegative and sum up to the
number of maximal consistent subset (via the efficiency property), we conclude that an FPRAS for $\shapi(D,\Delta,\Imc,f)$ implies an
FPRAS for the number of maximal matchings. To the best of our
knowledge, existence of the latter is an open problem. The same holds for any FD set $\Delta'$ that is not equivalent to an FD set with an
lhs chain, since there is a fact-wise reduction from $\Delta$ to such
$\Delta'$~\cite{DBLP:conf/pods/LivshitsK17}. Hence, the following remains unknown.
\begin{openpb}
	\label{open:shapi2}
 For any FD set $\Delta$ that has no lhs chain, is there a multiplicative FPRAS for $\shapi(\Delta,\Imc)$?
\end{openpb}
Actually, \Cref{open:pqe-to-shap} is solved in one special case.  It has been recently shown that the problem of counting the maximal consistent subsets for the FD set $\{A\rightarrow B,C\rightarrow D\}$ over $R(A,B,C,D)$ does not admit an FPRAS, unless 
$\mbox{NP}=\mbox{RP}$~\cite{DBLP:conf/pods/CalauttiLPS22a}. Therefore, we conclude that the same holds for the problem of computing  $\shapi(D,\{A\rightarrow B,C\rightarrow D\},\Imc,f)$.

As said earlier, \Cref{table:complexity} summarizes the complexity results for both the exact and approximate variants of the problems we studied here.

\subsection{Contribution to Cleaning Actions}
We presented the theoretical results of Livshits and Kimelfeld~\cite{DBLP:journals/lmcs/LivshitsK22} on the contribution of facts to inconsistency. 
The application of the Shapley value to machinery for data repairing has been proposed and studied by Deutch et al.~\cite{DBLP:conf/cikm/DeutchFGS21} from a different angle and a practical treatment. They aimed for explaining the individual actions of a data-repairing system, and particularly its decision to change a given entry in the database. They used the Shapley value to quantify the contribution of \e{relation cells} and the \e{ICs} to the decision to change a value. Hence, cells and ICs are the players in the coalitional game of causing the repairing algorithm to change a given cell. While their solutions do not have theoretical guarantees, they give heuristics that apply for a general setting: ICs can be any set of \e{denial constraints} (that widely generalize FDs) and repairing can be done by an arbitrary \e{black-box} algorithm.

\section{Concluding Remarks}\label{sec:conclusions}
We presented recent work on the application of the Shapley value to database tasks. Our focus has been on the theoretical analysis of the complexity of the underlying computational problems, and we discussed mainly two of them: computing the contribution of a fact to a query answer, and computing its contribution to  inconsistency. While there has been considerable research effort on the practical realization of the first challenge~\cite{deutch2022computing,arad2022learnshapley,DBLP:conf/sigmod/DavidsonDFKKM22,dosso2022credit}, the practical aspects of the contribution to inconsistency are yet to be explored (with the exception of Deutch et al.\cite{DBLP:conf/cikm/DeutchFGS21} that, as we explained, studied a problem that is related but different). Hence, many problems remain for future research, both on the theoretical side (e.g., the open problems we listed throughout the manuscript) and the practical one.

Notably, the framework for applying the Shapley
value to measuring contribution to query answers~\cite{lmcs} has been applied for proposing new bibliometric metrics~\cite{dosso2022credit}. The authors did not
use the algorithms proposed by Deutch et al.~\cite{deutch2022computing} (which are the most efficient to date, to the best of our knowledge), but instead computed the values directly from the definition. In this context, we believe that it is important to facilitate the use of existing algorithms by incorporating them into Postgres extensions such as ProvSQL~\cite{senellart2018provsql}.

\paragraph{\bf Acknowledgements.} 
Much of the work that we described in this survey was supported by the Israel Science Foundation (ISF), Grant 768/19, and the German Research Foundation (DFG) Project 412400621 (DIP program). L. Bertossi has been funded by ANID - Millennium Science
Initiative Program- Code ICN17002.

\balance

\bibliographystyle{abbrv} 

{
\small
\bibliography{references}

\begin{thebibliography}{10}

\bibitem{DBLP:conf/icdt/Amarilli23}
A.~Amarilli.
\newblock Uniform reliability for unbounded homomorphism-closed graph queries.
\newblock In {\em {ICDT}}, volume 255 of {\em LIPIcs}, pages 14:1--14:17, 2023.

\bibitem{arad2022learnshapley}
D.~Arad, D.~Deutch, and N.~Frost.
\newblock {LearnShapley}: Learning to predict rankings of facts contribution
  based on query logs.
\newblock In {\em CIKM}, pages 4788--4792, 2022.

\bibitem{AAAI21}
M.~Arenas, P.~Barcel{\'{o}}, L.~Bertossi, and M.~Monet.
\newblock The tractability of {SHAP}-score-based explanations for
  classification over deterministic and decomposable boolean circuits.
\newblock In {\em {AAAI}}, pages 6670--6678, 2021.

\bibitem{arenas2021complexity}
M.~Arenas, P.~Barcel\'o, L.~Bertossi, and M.~Monet.
\newblock On the complexity of {SHAP}-score-based explanations: Tractability
  via knowledge compilation and non-approximability results.
\newblock {\em Journal of Machine Learning Research}, 24(63):1--58, 2023.

\bibitem{DBLP:conf/pods/ArenasBC99}
M.~Arenas, L.~E. Bertossi, and J.~Chomicki.
\newblock Consistent query answers in inconsistent databases.
\newblock In {\em {PODS}}, pages 68--79. {ACM} Press, 1999.

\bibitem{DBLP:journals/vldb/BenjellounSHTW08}
O.~Benjelloun, A.~D. Sarma, A.~Y. Halevy, M.~Theobald, and J.~Widom.
\newblock Databases with uncertainty and lineage.
\newblock {\em {VLDB} J.}, 17(2):243--264, 2008.

\bibitem{DBLP:conf/pods/Bertossi19}
L.~Bertossi.
\newblock Database repairs and consistent query answering: Origins and further
  developments.
\newblock In D.~Suciu, S.~Skritek, and C.~Koch, editors, {\em PODS}, pages
  48--58. {ACM}, 2019.

\bibitem{DBLP:conf/lpnmr/Bertossi19}
L.~Bertossi.
\newblock Repair-based degrees of database inconsistency.
\newblock In {\em {LPNMR}}, volume 11481 of {\em LNCS}, pages 195--209.
  Springer, 2019.

\bibitem{DBLP:journals/kais/Bertossi21}
L.~Bertossi.
\newblock Specifying and computing causes for query answers in databases via
  database repairs and repair-programs.
\newblock {\em Knowl. Inf. Syst.}, 63(1):199--231, 2021.

\bibitem{LeoRW22}
L.~Bertossi.
\newblock Attribution-scores and causal counterfactuals as explanations in
  artificial intelligence.
\newblock In {\em Bertossi, L., Xiao, G. (eds.) Reasoning Web. Causality,
  Explanations and Declarative Knowledge. Springer LNCS 13759}, pages 1--23,
  2023.

\bibitem{deem}
L.~Bertossi, J.~Li, M.~Schleich, D.~Suciu, and Z.~Vagena.
\newblock Causality-based explanation of classification outcomes.
\newblock In {\em DEEM@SIGMOD}, pages 6:1--6:10. {ACM}, 2020.

\bibitem{DBLP:journals/ijar/BertossiS17}
L.~Bertossi and B.~Salimi.
\newblock Causes for query answers from databases: Datalog abduction,
  view-updates, and integrity constraints.
\newblock {\em Int. J. Approx. Reason.}, 90:226--252, 2017.

\bibitem{DBLP:journals/mst/BertossiS17}
L.~Bertossi and B.~Salimi.
\newblock From causes for database queries to repairs and model-based diagnosis
  and back.
\newblock {\em Theory Comput. Syst.}, 61(1):191--232, 2017.

\bibitem{DBLP:journals/sigmod/BunemanT18}
P.~Buneman and W.~Tan.
\newblock Data provenance: What next?
\newblock {\em {SIGMOD} Rec.}, 47(3):5--16, 2018.

\bibitem{Burkart}
N.~Burkart and M.~F. Huber.
\newblock A survey on the explainability of supervised machine learning.
\newblock {\em J. Artif. Intell. Res.}, 70:245--317, 2021.

\bibitem{DBLP:conf/pods/CalauttiLPS22a}
M.~Calautti, E.~Livshits, A.~Pieris, and M.~Schneider.
\newblock Counting database repairs entailing a query: The case of functional
  dependencies.
\newblock In {\em {PODS}}, pages 403--412. {ACM}, 2022.

\bibitem{Chockler04}
H.~Chockler and J.~Y. Halpern.
\newblock Responsibility and blame: {A} structural-model approach.
\newblock {\em J. Artif. Intell. Res.}, 22:93--115, 2004.

\bibitem{DBLP:conf/atal/CholvyPRT15}
L.~Cholvy, L.~Perrussel, W.~Raynaut, and J.-M. Th\'evenin.
\newblock Towards consistency-based reliability assessment.
\newblock In {\em {AAMAS}}, pages 1643--1644. {ACM}, 2015.

\bibitem{dalvi2013dichotomy}
N.~Dalvi and D.~Suciu.
\newblock The dichotomy of probabilistic inference for unions of conjunctive
  queries.
\newblock {\em Journal of the ACM (JACM)}, 59(6):1--87, 2013.

\bibitem{DBLP:journals/cacm/DalviRS09}
N.~N. Dalvi, C.~R{\'{e}}, and D.~Suciu.
\newblock Probabilistic databases: Diamonds in the dirt.
\newblock {\em Commun. {ACM}}, 52(7):86--94, 2009.

\bibitem{darwiche2004new}
A.~Darwiche.
\newblock New advances in compiling {CNF} to decomposable negation normal form.
\newblock In {\em Proceedings of ECAI}, pages 328--332. Citeseer, 2004.

\bibitem{DBLP:conf/sigmod/DavidsonDFKKM22}
S.~B. Davidson, D.~Deutch, N.~Frost, B.~Kimelfeld, O.~Koren, and M.~Monet.
\newblock {ShapGraph}: An holistic view of explanations through provenance
  graphs and {Shapley} values.
\newblock In {\em {SIGMOD} Conference}, pages 2373--2376. {ACM}, 2022.

\bibitem{DBLP:conf/cikm/DeutchFGS21}
D.~Deutch, N.~Frost, A.~Gilad, and O.~Sheffer.
\newblock Explanations for data repair through {Shapley} values.
\newblock In {\em {CIKM}}, pages 362--371. {ACM}, 2021.

\bibitem{deutch2022computing}
D.~Deutch, N.~Frost, B.~Kimelfeld, and M.~Monet.
\newblock Computing the {Shapley} value of facts in query answering.
\newblock In {\em SIGMOD}, pages 1570--1583, 2022.

\bibitem{dosso2022credit}
D.~Dosso, S.~B. Davidson, and G.~Silvello.
\newblock Credit distribution in relational scientific databases.
\newblock {\em Information Systems}, 109:102060, 2022.

\bibitem{DBLP:journals/jacm/GoldreichGR98}
O.~Goldreich, S.~Goldwasser, and D.~Ron.
\newblock Property testing and its connection to learning and approximation.
\newblock {\em J. {ACM}}, 45(4):653--750, 1998.

\bibitem{DBLP:journals/jiis/GrantH06}
J.~Grant and A.~Hunter.
\newblock Measuring inconsistency in knowledgebases.
\newblock {\em J. Intell. Inf. Syst.}, 27(2):159--184, 2006.

\bibitem{DBLP:conf/ecsqaru/GrantH11}
J.~Grant and A.~Hunter.
\newblock Measuring consistency gain and information loss in stepwise
  inconsistency resolution.
\newblock In {\em ECSQARU}, volume 6717 of {\em LNCS}, pages 362--373.
  Springer, 2011.

\bibitem{DBLP:conf/ecsqaru/GrantH13}
J.~Grant and A.~Hunter.
\newblock Distance-based measures of inconsistency.
\newblock In {\em {ECSQARU}}, volume 7958 of {\em LNCS}, pages 230--241.
  Springer, 2013.

\bibitem{DBLP:journals/ijar/GrantH17}
J.~Grant and A.~Hunter.
\newblock Analysing inconsistent information using distance-based measures.
\newblock {\em Int. J. Approx. Reasoning}, 89:3--26, 2017.

\bibitem{DBLP:conf/pods/GreenT17}
T.~J. Green and V.~Tannen.
\newblock The semiring framework for database provenance.
\newblock In E.~Sallinger, J.~V. den Bussche, and F.~Geerts, editors, {\em
  PODS}, pages 93--99. {ACM}, 2017.

\bibitem{fosca}
R.~Guidotti, A.~Monreale, S.~Ruggieri, F.~Turini, F.~Giannotti, and
  D.~Pedreschi.
\newblock A survey of methods for explaining black box models.
\newblock {\em {ACM} Comput. Surv.}, 51(5):93:1--93:42, 2019.

\bibitem{halpern}
J.~Y. Halpern.
\newblock {\em Actual Causality}.
\newblock {MIT} Press, 2016.

\bibitem{HP05}
J.~Y. Halpern and J.~Pearl.
\newblock Causes and explanations: A structural-model approach. part i: Causes.
\newblock {\em British Journal for the Philosophy of Science}, 56(4):843--887,
  2005.

\bibitem{Halpern2005-HALCAE-2}
J.~Y. Halpern and J.~Pearl.
\newblock Causes and explanations: A structural-model approach. part ii:
  Explanations.
\newblock {\em British Journal for the Philosophy of Science}, 56(4):889--911,
  2005.

\bibitem{DBLP:conf/kr/HunterK06}
A.~Hunter and S.~Konieczny.
\newblock {Shapley} inconsistency values.
\newblock In {\em {KR}}, pages 249--259. {AAAI} Press, 2006.

\bibitem{DBLP:conf/kr/HunterK08}
A.~Hunter and S.~Konieczny.
\newblock Measuring inconsistency through minimal inconsistent sets.
\newblock In {\em KR}, pages 358--366. {AAAI} Press, 2008.

\bibitem{DBLP:journals/ai/HunterK10}
A.~Hunter and S.~Konieczny.
\newblock On the measure of conflicts: {Shapley} inconsistency values.
\newblock {\em Artif. Intell.}, 174(14):1007--1026, 2010.

\bibitem{imbens}
G.~W. Imbens and D.~B. Rubin.
\newblock {\em Causal Inference for Statistics, Social, and Biomedical
  Sciences: An Introduction}.
\newblock Cambridge University Press, 2015.

\bibitem{khalil2022complexity}
M.~Khalil and B.~Kimelfeld.
\newblock The complexity of the {Shapley} value for regular path queries.
\newblock {\em arXiv preprint arXiv:2212.07720}, 2022.

\bibitem{DBLP:conf/pods/KimelfeldVW11}
B.~Kimelfeld, J.~Vondr{\'{a}}k, and R.~Williams.
\newblock Maximizing conjunctive views in deletion propagation.
\newblock In {\em {PODS}}, pages 187--198. {ACM}, 2011.

\bibitem{DBLP:conf/ijcai/KoniecznyLM03}
S.~Konieczny, J.~Lang, and P.~Marquis.
\newblock Quantifying information and contradiction in propositional logic
  through test actions.
\newblock In {\em {IJCAI}}, pages 106--111. Morgan Kaufmann, 2003.

\bibitem{lmcs}
E.~Livshits, L.~Bertossi, B.~Kimelfeld, and M.~Sebag.
\newblock The {Shapley} value of tuples in query answering.
\newblock {\em Log. Methods Comput. Sci.}, 17(3), 2021.

\bibitem{DBLP:conf/pods/LivshitsK17}
E.~Livshits and B.~Kimelfeld.
\newblock Counting and enumerating (preferred) database repairs.
\newblock In {\em {PODS}}, pages 289--301. {ACM}, 2017.

\bibitem{DBLP:journals/lmcs/LivshitsK22}
E.~Livshits and B.~Kimelfeld.
\newblock The {Shapley} value of inconsistency measures for functional
  dependencies.
\newblock {\em Log. Methods Comput. Sci.}, 18(2), 2022.

\bibitem{DBLP:journals/tods/LivshitsKR20}
E.~Livshits, B.~Kimelfeld, and S.~Roy.
\newblock Computing optimal repairs for functional dependencies.
\newblock {\em {ACM} Trans. Database Syst.}, 45(1):4: 1--4: 46, 2020.

\bibitem{DBLP:journals/jcss/LivshitsKW21}
E.~Livshits, B.~Kimelfeld, and J.~Wijsen.
\newblock Counting subset repairs with functional dependencies.
\newblock {\em J. Comput. Syst. Sci.}, 117:154--164, 2021.

\bibitem{DBLP:conf/sigmod/LivshitsKTIKR21}
E.~Livshits, R.~Kochirgan, S.~Tsur, I.~F. Ilyas, B.~Kimelfeld, and S.~Roy.
\newblock Properties of inconsistency measures for databases.
\newblock In {\em {SIGMOD}}, pages 1182--1194. {ACM}, 2021.

\bibitem{lundberg20}
S.~M. Lundberg, G.~G. Erion, H.~Chen, A.~J. DeGrave, J.~M. Prutkin, B.~Nair,
  R.~Katz, J.~Himmelfarb, N.~Bansal, and S.~Lee.
\newblock From local explanations to global understanding with explainable {AI}
  for trees.
\newblock {\em Nat. Mach. Intell.}, 2(1):56--67, 2020.

\bibitem{lund17}
S.~M. Lundberg and S.~Lee.
\newblock A unified approach to interpreting model predictions.
\newblock In {\em {NIPS}}, pages 4765--4774, 2017.

\bibitem{DBLP:journals/ton/MaCLMR10}
R.~T.~B. Ma, D.~Chiu, J.~C. Lui, V.~Misra, and D.~Rubenstein.
\newblock Internet economics: The use of {Shapley} value for {ISP} settlement.
\newblock {\em {IEEE/ACM} Trans. Netw.}, 18(3):775--787, 2010.

\bibitem{DBLP:journals/pvldb/MeliouGMS11}
A.~Meliou, W.~Gatterbauer, K.~F. Moore, and D.~Suciu.
\newblock The complexity of causality and responsibility for query answers and
  non-answers.
\newblock {\em Proc. {VLDB} Endow.}, 4(1):34--45, 2010.

\bibitem{minh}
D.~Minh, H.~Wang, Y.~Li, and T.~Nguyen.
\newblock Explainable artificial intelligence: a comprehensive review.
\newblock {\em Artificial Intelligence Review}, 55, 11 2021.

\bibitem{molnar}
C.~Molnar.
\newblock {\em Interpretable Machine Learning}.
\newblock https://christophm.github.io/interpretable-ml-book/, 2019.

\bibitem{monet2020solving}
M.~Monet.
\newblock Solving a special case of the intensional vs extensional conjecture
  in probabilistic databases.
\newblock In {\em Proceedings of PODS}, pages 149--163, 2020.

\bibitem{moretti2007class}
S.~Moretti, F.~Patrone, and S.~Bonassi.
\newblock The class of microarray games and the relevance index for genes.
\newblock {\em Top}, 15(2):256--280, 2007.

\bibitem{DBLP:journals/tase/NarayanamN11}
R.~Narayanam and Y.~Narahari.
\newblock A {Shapley} value-based approach to discover influential nodes in
  social networks.
\newblock {\em {IEEE} Trans Autom. Sci. Eng.}, 8(1):130--147, 2011.

\bibitem{pearl}
J.~Pearl.
\newblock {\em Causality: Models, Reasoning and Inference}.
\newblock Cambridge University Press, 2nd edition, 2009.

\bibitem{DBLP:conf/pods/ReshefKL20}
A.~Reshef, B.~Kimelfeld, and E.~Livshits.
\newblock The impact of negation on the complexity of the {Shapley} value in
  conjunctive queries.
\newblock In {\em {PODS}}, pages 285--297. {ACM}, 2020.

\bibitem{roth}
A.~E. Roth, editor.
\newblock {\em The {Shapley} value : essays in honor of Lloyd S. Shapley}.
\newblock Cambridge University Press, 1988.

\bibitem{tapp}
B.~Salimi, L.~Bertossi, D.~Suciu, and G.~V. den Broeck.
\newblock Quantifying causal effects on query answering in databases.
\newblock In {\em TaPP}. {USENIX} Association, 2016.

\bibitem{senellart2018provsql}
P.~Senellart, L.~Jachiet, S.~Maniu, and Y.~Ramusat.
\newblock {ProvSQL}: Provenance and probability management in {PostgreSQL}.
\newblock {\em Proc. {VLDB} Endow.}, 11(12):2034--2037, 2018.

\bibitem{shapley:book1952}
L.~S. Shapley.
\newblock A value for n-person games.
\newblock In H.~W. Kuhn and A.~W. Tucker, editors, {\em Contributions to the
  Theory of Games II}, pages 307--317. Princeton University Press, Princeton,
  1953.

\bibitem{struss}
P.~Struss.
\newblock Model-based problem solving.
\newblock In {\em Handbook of Knowledge Representation}, 2008.

\bibitem{suciu}
D.~Suciu, D.~Olteanu, C.~R{\'{e}}, and C.~Koch.
\newblock {\em Probabilistic Databases}.
\newblock Synthesis Lectures on Data Management. Morgan {\&} Claypool
  Publishers, 2011.

\bibitem{DBLP:journals/ki/Thimm17}
M.~Thimm.
\newblock On the compliance of rationality postulates for inconsistency
  measures: {A} more or less complete picture.
\newblock {\em {KI}}, 31(1):31--39, 2017.

\bibitem{DBLP:journals/snam/CampenHHL18}
T.~van Campen, H.~Hamers, B.~Husslage, and R.~Lindelauf.
\newblock A new approximation method for the {Shapley} value applied to the
  {WTC} 9/11 terrorist attack.
\newblock {\em Soc. Netw. Anal. Min.}, 8(1):3:1--3:12, 2018.

\bibitem{guyAAAI21}
G.~{Van den Broeck}, A.~Lykov, M.~Schleich, and D.~Suciu.
\newblock On the tractability of {SHAP} explanations.
\newblock {\em J. Artif. Intell. Res.}, 74:851--886, 2022.

\bibitem{vardi1982complexity}
M.~Y. Vardi.
\newblock The complexity of relational query languages.
\newblock In {\em STOC}, pages 137--146. ACM, 1982.

\end{thebibliography}
}

\end{document}